\documentclass[fleqn,usenatbib]{mnras}

\usepackage{newtxtext,newtxmath}
\usepackage[T1]{fontenc}
\usepackage{ae,aecompl}

\usepackage{graphicx}	% Including figure files
\usepackage{amsmath}	% Advanced maths commands
\usepackage{amssymb}	% Extra maths symbols

\title[On the evolutionary status and pulsations of the BLAPs]{On the evolutionary status and pulsations of the recently discovered Blue Large-Amplitude Pulsators (BLAPs)}

\author[Romero et al.]{Alejandra D. Romero$^{1}$\thanks{E-mail: alejandra.romero@ufrgs.br},
A. H. C\'orsico$^{2,3}$, L. G. Althaus$^{2,3}$, I. Pelisoli$^{1}$
\newauthor and S. O. Kepler$^{1}$ \\
$^{1}$Physics Institute, Universidade Federal do Rio Grande do Sul, Av. Bento Gon\c{c}alves 9500, Brazil\\
$^{2}$Facultad de Ciencias Astron\'omicas y Geof\'isicas, Universidad Nacional de La Plata, La Plata 1900, Argentina\\
$^{3}$CONICET, Consejo Nacional de Investigaciones Cientif\'icas y T\'ecnicas, Argentina
}

\date{Accepted XXX. Received YYY; in original form ZZZ}

\pubyear{2018}

\begin{document}
\label{firstpage}
\pagerange{\pageref{firstpage}--\pageref{lastpage}}
\maketitle

% Abstract of the paper   200 words max.
\begin{abstract}
The Blue Large-Amplitude Pulsators (BLAPs) constitute a new class of pulsating stars. They are hot stars with effective temperatures of $\sim 30\, 000$ K and surface gravities of $\log g \sim 4.9$, that pulsate with periods in the range $20-40$ min. Until now, their origin and evolutionary state, as well as the nature of their pulsations, were not been unveiled. In this paper, we propose that the BLAPs are the hot counterpart of the already known pulsating pre-Extremely Low Mass (pre-ELM) white dwarf (WD) stars, that are He-core low-mass stars resulting from interacting binary evolution. Using fully evolutionary sequences, we show that the BLAPs are well represented by pre-ELM WD models with high effective temperature and stellar masses $\sim 0.34 M_{\sun}$. From the analysis of their pulsational properties, we find that the observed variabilities can be explained by high--order nonradial $g-$mode pulsations or, in the case of the shortest periods, also by low--order radial modes, including the fundamental radial mode. The theoretical modes with periods in the observed range are unstable due to the $\kappa$ mechanism associated to the $Z$-bump in the opacity at $\log T \sim 5.25$. 
\end{abstract}


%%%%%%%%%%%%%%%%%%%%%%%%%%%%%%%%%%%%%%%%%%%%%%%%%%

%%%%%%%%%%%%%%%%% BODY OF PAPER %%%%%%%%%%%%%%%%%%

\section{Introduction}

Across the Hertzprung-Russell (H-R) diagram, there are several different classes of pulsating stars, from the main sequence (MS) to the white dwarf (WD) cooling curve. Pulsations give valuable information about the inner structure of the pulsating stars, otherwise invisible from direct observations. Recently, \citet{2017NatAs...1E.166P} reported the discovery of a group of long-period variable stars named as Blue Large-Amplitude Pulsators (BLAPs), as a result of a long time photometric study of the OGLE. These objects show regular brightness variations with periods in the range $20-40$ min and amplitudes of $0.2-0.4$ mag in the optical. The light curves have a characteristic sawtooth shape, similar to the shape of classical Cepheids and RR Lyrae-type stars that pulsate in the fundamental radial mode. The BLAP stars are bluer than MS stars observed in the same fields, indicating that they are hot stars. Follow-up spectroscopy confirms a high surface temperature of about $30\, 000$ K \citep{2017NatAs...1E.166P}. 

The BLAP stars show surface gravities similar to MS pulsators, as $\delta$ Scuti stars, however, their effective temperatures are much hotter than those of $\delta$ Scuti stars (Fig. \ref{evolution}). Compact pusators, in particular variable sub-dwarf stars (V361 Hya and V1093 Her), show effective temperatures similar to the BLAPs, but their surface gravities are higher (see Fig. \ref{evolution}) and also the characteristic periods are short, as compared to the periods observed in BLAPs. Finally, a class of stars recently uncovered is the pre-extremely low mass (pre-ELM) WDs, that are the evolutionary precursors of the Extremely Low Mass (ELM) WD stars. From evolutionary computations, it is known that the pre-ELM WDs
are in the high-luminosity stage induced by residual H burning in He-core degenerate low-mass stars \citep[see e.g.][]{2013A&A...557A..19A, 2014A&A...571A..45I}. At this stage, the evolutionary sequences can reach effective temperatures as high as $100\,000$ K, depending on the stellar mass \citep{2013A&A...557A..19A}. In addition, the pre-ELM WD class have five pulsating members known to date \citep[the so called pre-ELMVs, see, e.g.,][]{2013Natur.498..463M, 2014MNRAS.444..208M, 2016ApJ...822L..27G}, which show surface gravities and effective temperatures of $\log g \sim 4.9$ and $8600  \lesssim T_{\rm eff} \lesssim 11\,900$ K, respectively.

In this Letter, we propose that BLAP stars are hot pulsating pre-ELM WD stars with masses $\sim 0.30-0.40 M_{\sun}$. We present the evolutionary scenario in Section \ref{zwei}. Adiabatic and non-adiabatic pulsation analyses are presented in Sections \ref{drei} and \ref{vier}, respectively. Concluding remarks are presented in Section \ref{conclusion}.

%Long term photometric observations show that the variable stars are very stable over time. Derived rates of period change are of the order of $10^{-7}$ per year and, in most cases, they are positive.

\section{Evolutionary origin}\label{zwei}

The evolutionary origin for the BLAP stars was preliminarily explored by \citet{2017NatAs...1E.166P} using simple models, but it still remains a mystery. These stars are too hot to be $\delta$ Scuti stars and their surface gravity is not compatible with the sub-dwarf compact pulsators. However, they show similar surface gravity than the pre-ELMVs and higher effective temperatures, implying that they could be massive ($M_{\star} \gtrsim 0.30 M_{\sun}$) pre-ELM WDs evolving during the high-effective temperature, high-luminosity stage induced by residual H-shell burning, characteristic of low-mass WD evolution \citep{2013A&A...557A..19A, 2014A&A...571A..45I}.

\begin{figure}
\includegraphics[width=\columnwidth]{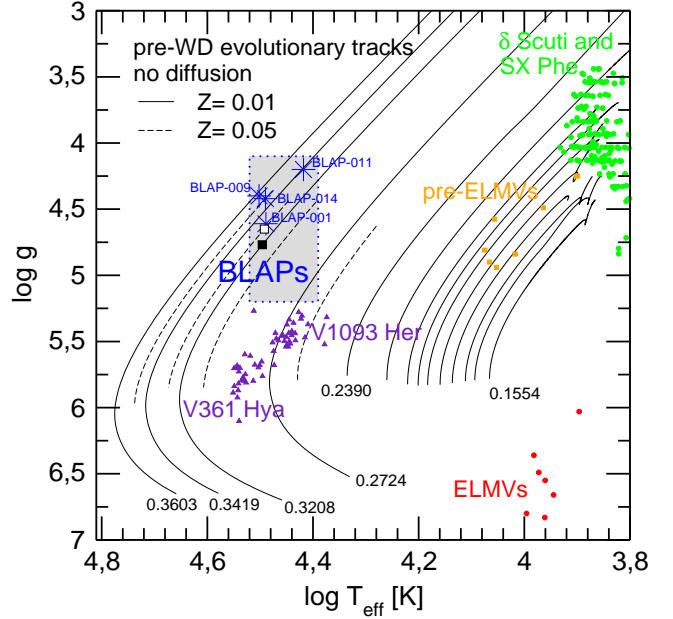}
\caption{$\log T_{\rm eff}- \log g$ diagram showing the location of  the
BLAP stars (shaded rectangle area), along with other classes of
pulsating stars: ELMVs (red dots), pre-ELMVs (orange dots), pulsating sdBs
(V361 Hya and V1093 Her; violet triangles) and $\delta$ Sct/SX Phe
stars (green dots). Solid black lines correspond to low-mass
He-core pre-WD evolutionary tracks computed neglecting element
diffusion and $Z= 0.01$. Numbers  correspond  to the  stellar mass  of
some sequences. We also include portions of evolutionary tracks
corresponding to $Z= 0.05$ for some stellar masses.  Blue star symbols  
indicate the location of four BLAP stars with measured  
atmospheric parameters: BLAP-001 (the prototype star), BLAP-009, 
BLAP-011, and BLAP-014 (see Table \ref{tab:compare}). The symbols on
the evolutionary tracks of $M_{\star}= 0.3208$ and $Z= 0.01$ (black
square) and $M_{\star}= 0.3419$ and $Z= 0.05$ (white square) indicate
the location of two template models analyzed in the text.}
\label{evolution}
\end{figure}

To test this hypothesis, we present in Fig, \ref{evolution} the position of the BLAPs in the $\log T_{\rm eff}- \log g$ diagram, along with other classes of pulsating stars. The figure includes low-mass He-core pre-ELM WD evolutionary tracks computed neglecting element diffusion and metallicity $Z= 0.01$ (solid lines), with initial models extracted from \citet{2013A&A...557A..19A}. Additional sequences with an enhanced metallicity of $Z= 0.05$ (dashed lines) were computed specifically for this work. 

We neglect element diffusion and rotational mixing to reproduce the atmosphere composition observed in the BLAPs, containing a mixture of H and He. As it was shown by \citet{2016A&A...595A..35I}, rotational mixing and gravitational settling compete with each other to determine the composition of the atmosphere in pre-ELM WDs. These authors found that at the initial stages of the shell-flashes, rotational mixing dominates leading to a mixed atmosphere composition. As the surface gravity increases, gravitational settling overcomes rotational mixing, and only by the beginning of the last flash gravitational settling completely dominates.

Note that the region where the BLAPs are found is perfectly covered by evolutionary tracks characterized by stellar masses between $\sim 0.27 - 0.37 M_{\sun}$. At this stage, the stars have an extended envelope, compatible with long-period pulsators. Then, we can conclude that the BLAPs are the high-effective temperature, high-mass counterparts of the previously known pre-ELMVs stars. Finally, employing our evolutionary tracks, we estimate the stellar mass for the four BLAPs with measured atmospheric parameters assuming the two metallicity values considered in our computations. The results are listed in Table \ref{tab:compare}. Note that the values of the stellar mass from the evolutionary sequences are compatible with the spectroscopic masses estimated by \citet{2017NatAs...1E.166P} using simplified shell H-burning models.

\section{Adiabatic pulsations}\label{drei}

Detection of pulsations in the BLAP stars opens an unprecedented opportunity of probing the internal structure of this new class of variable stars. Following the results obtained in Section \ref{zwei}, here we use the pulsational properties of our theoretical models to confirm the evolutionary origin of the new pulsators. We carried out a detailed adiabatic radial ($\ell= 0$) and non-radial ($\ell=1,2$) $p-$ and $g-$mode pulsation analysis. The pulsation computations were performed with the adiabatic version of the pulsation code {\tt LP-PUL} that is described in \citet{2006A&A...454..863C} which is coupled to the {\tt LPCODE} evolutionary code. Details on pulsation computations for variable ELM and pre-ELM WD stars can be found in \citet{2014A&A...569A.106C, 2016A&A...585A...1C, 2016A&A...588A..74C}.

\begin{table}
	\centering
	\caption{Stellar mass (in solar units) determinations for the four stars with estimated atmospheric parameters \citep{2017NatAs...1E.166P} (columns 2 and 3), considering two metallicities (columns 4 and 5).  }
	\label{tab:compare}
	\begin{tabular}{lcccc} % four columns, alignment for each
		\hline
		star & $T_{\rm eff}$ & $\log g$ & $M_{\star}/M_{\sun}$ & $M_{\star}/M_{\sun}$\\
   		& & & $(Z= 0.01)$ & $(Z= 0.05)$\\
		\hline
BLAP-001 & $30\, 800 \pm 500$  & $4.61 \pm 0.07$ & 0.3320 &   0.3439\\
BLAP-009 & $31\, 800 \pm 1400$ & $4.40 \pm 0.18$ & 0.3571 &   0.3666\\
BLAP-011 & $26\, 200 \pm 2900$ & $4.20 \pm 0.29$ & 0.3419 &   0.3516\\
BLAP-014 & $30\, 900 \pm 2100$ & $4.42 \pm 0.26$ & 0.3504 &   0.3603\\
		\hline
	\end{tabular}
\end{table}

\begin{figure}
\includegraphics[width=\columnwidth]{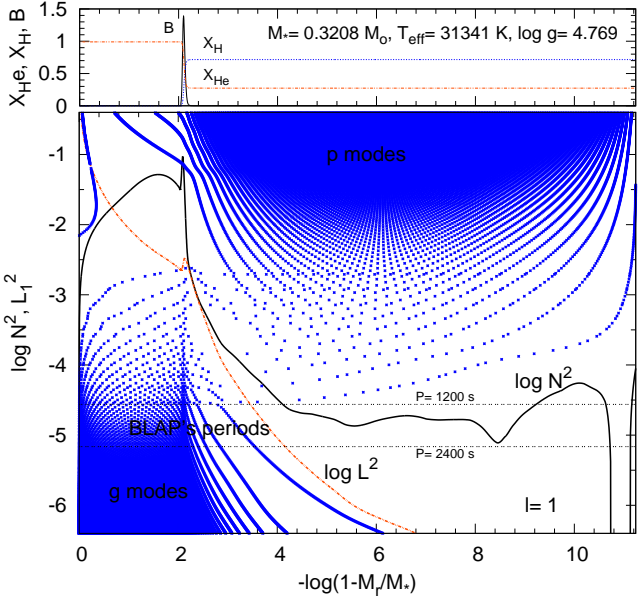}
\caption{Internal chemical profiles of He and H and the Ledoux term B
(upper panel) and the propagation diagram ---the run of the logarithm
of the squared critical frequencies ($N^2$, $L^2_{\ell= 1}$), lower
panel--- in terms of the outer mass fraction coordinate, corresponding
to the pre-ELM WD template model of $M_{\star}= 0.3208 M_{\sun}$ and
$T_{\rm eff} \sim 31\,300$ K marked in Fig. \ref{evolution} with a black square. In the lower panel, tiny star symbols (in blue) correspond to the
spatial location of the nodes of the radial eigenfunction of dipole
($\ell= 1$) $g$ and $p$ modes. For squared frequencies in the range
$-2.5 \gtrsim \log \sigma^2 \gtrsim -4.5$ the modes behave like mixed
$p-g$ modes. The (squared) frequency interval corresponding to the
modes detected in BLAP stars (with periods between $\sim 1200$ s and $\sim 2400$ s) is enclosed with two horizontal dotted lines.  }
\label{fig02}
\end{figure}

To illustrate the adiabatic pulsation properties of the pre-ELM WD models representative of the BLAP stars, we chose a template model, marked with a black square in Fig. \ref{evolution}, characterized by $M_{\star}= 0.3208 M_{\sun}$, $Z= 0.01$ and $T_{\rm eff} \sim 31300$ K. In Fig. \ref{fig02} we present the internal chemical profile for H and He and the Ledoux term $B$, that is crucial for the computation of the Brunt-V\"ais\"ala frequency (upper panel), along with the propagation diagram in terms of the outer mass fraction (lower panel). The period range for our template model extends from 10 s to $\sim 10\, 000$ s. The sharp chemical He/H transition region leaves notorious signatures on the run of the square of the critical frequencies, in particular for the Brunt-V\"ais\"ala frequency $N$, at $\log(1-M_r/M_{\star}) \sim -2$ ($r/R_{\star}= 0.069$), as can be seen from Fig.  \ref{fig02}. The period range observed in BLAP stars is indicated with two horizontal dotted lines and can be associated to high--order nonradial $g-$ modes, with radial order in the range $25 \lesssim k  \lesssim 50$ for modes with $\ell= 1$. A similar result is obtained if we consider $\ell= 2$ modes (not shown). On the other hand, low--order radial ($\ell= 0$) modes, including the fundamental mode with $k=0$, show periods $\sim 1200$ s, consistent with the shortest periods observed in the BLAP stars.

Finally, \citet{2017NatAs...1E.166P} determined a rate of period change for 11 objects. As a result, they obtained a value of $\dot{\Pi}\sim 10^{-7}$/yr, implying that the objects are evolving on a nuclear timescale. For the template model, $\dot{\Pi}\sim \pm 10^{-7}- 10^{-5}$/yr in the period range observed in the BLAPs stars, in good agreement with the observed value. In addition, our models reproduce the negative $\dot{\Pi}$ values observed for two objects, which cannot be easily accounted for by other evolutionary scenarios.

\section{Non-adiabatic pulsations}\label{vier}

In this Section, we analyze the pulsation stability properties of models representative of the BLAP stars. We compute non-adiabatic pulsations for pre-ELM WD stars models with stellar mass $0.2724, 0.3208, 0.3414$ and $0.3630 M_{\sun}$ and effective temperature in the range of $25\, 000 - 36\, 000$ K, covering the region where the BLAP stars are found. The non-adiabatic pulsation computations were carried out employing the non-adiabatic version of the {\tt LP-PUL} pulsation code, described in \citet{2006A&A...458..259C}.
A detailed analysis of the dependence of the stability pulsational properties of our models for BLAP stars with the stellar parameters ($M_{\star}, Z$, prescription of convection, etc) is out of the scope of this work, and will be addressed in a future paper.

\begin{figure}
\includegraphics[width=\columnwidth]{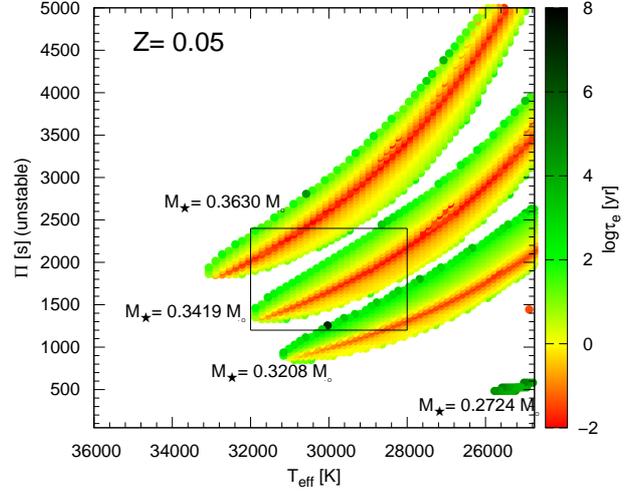}
\caption{Periods of unstable  ($\ell= 1$) $g$ modes in terms of the
effective temperature, with the palette of colors (right scale)
indicating the value of the logarithm of the $e$-folding time (in
years), corresponding to  He-core pre-WD sequences with stellar masses
$0.2724, 0.3208, 0.3419$ and $0.3630 M_{\sun}$, and $Z= 0.05$. Note
that the $e$-folding times range from $\sim 10^{-2}$ to $10^2$ yr,
much shorter than the typical evolutionary timescales at that stage of
evolution. The rectangle corresponds to the $T_{\rm eff}$ interval
measured for BLAPs and the range of observed periods. Note that the observed
periodicities ($1200 \lesssim \Pi \lesssim 2400$ s) are well accounted
for by our theoretical computations if we consider a range of stellar
masses ($0.33 \lesssim M_{\star}/M_{\sun} \lesssim 0.36$).}
\label{fig03}
\end{figure}

In this work, we hypothesize that the driving mechanism that excites the pulsations in BLAPs is the $\kappa$ mechanism, caused by a bump in the opacity due to heavy elements ($Z$--bump). This mechanism is the responsible for the pulsation instabilities found in sub-dwarf B variable stars \citep[see e.g.][]{1997ApJ...483L.123C,2003ApJ...597..518F}, characterized by an effective temperature similar to those estimated for the BLAPs (see Fig. \ref{evolution}). In particular, we found that pulsation modes from pre-ELM WD models characterized with a solar metallicity $Z= 0.01$ are not pulsationally unstable for high effective temperatures. However, we noted that there is strong driving at the region of the model in which the $Z$ bump is located. Guided by that finding, and in order to have sufficient driving to globally destabilize the modes at high effective temperatures characteristic of the BLAP stars, we artificially increased the total metallicity of the models to $Z= 0.05$. This greatly enhanced the "height" of the $Z$ bump in the opacity, and we finally were able to find unstable modes. 

Note that an increase in the amplitude of the $Z$-bump in the opacity profiles can be achieved by considering radiative levitation, which can build a concentration of the Fe-peak elements in the inner convection region without increasing the overall metallicity of the model. 
Since our models are not including radiative levitation, we use the artificial tool of increasing the total metallicity to increase the amplitude of the bump in the opacity due to $Z$-elements.  The effect of an enhanced metallicity mimics the effects of the radiative levitation in producing a $Z$-bump in the opacity profile. In particular, the evolutionary timescales may change when different progenitor metallicities are considered, leading to different values of $\dot{\Pi}$. However, for our models with $Z=0.01$ and $Z=0.05$ the values for $\dot{\Pi}$ show the same order of magnitude. The impact on the evolutionary timescales when a locally enhanced $Z$ model is considered (such as one that incorporates radiative levitation) has not been studied in the literature yet, and it deserves to be explored in detail.

Fig. \ref{fig03} shows the periods of dipolar unstable $g$ modes in terms of the effective temperature for four stellar masses and $Z=0.05$. The values of the $e$-folding times (color pallet in Fig. \ref{fig03}) range from $\sim 10^{-2}$ to $10^2$ yr, much shorter than the typical evolutionary timescales at that stage of evolution. The period range observed in the BLAPs, represented in Fig. \ref{fig03} as a rectangle, is well accounted for by our theoretical computations for stellar masses between $0.3208$ and $0.3630 M_{\sun}$. 

\begin{figure}
\includegraphics[width=\columnwidth]{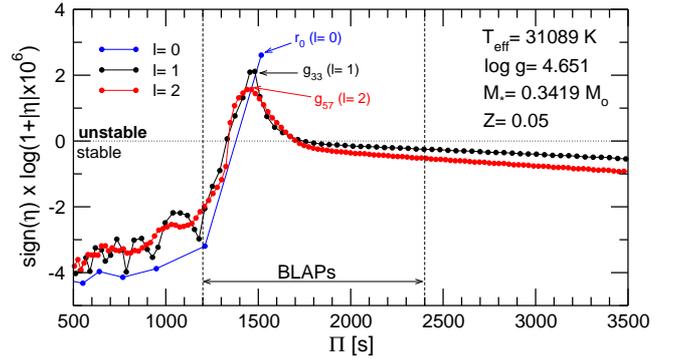}
\caption{Normalized growth rates $\eta$ ($\eta > 0$ implies
instability) for $\ell= 0$ (blue), $\ell= 1$ (black), and
$\ell= 2$ (red) modes in terms of the pulsation periods for
the $0.3419 M_{\sun}$ pre-WD template model at $T_{\rm eff}\sim 31100$
K marked  in Fig. \ref{evolution} with a white square. The  large  numerical  range spanned by $\eta$ is appropriately scaled for a better graphical
representation. Some specific modes ---the most unstable ones for each
considered $\ell$ value--- are labeled. The unstable $g$ mode
with $k= 33$ is analyzed in Fig. \ref{fig05}. The vertical dashed lines enclose the period interval observed in BLAP stars.}
\label{fig04}
\end{figure}

Figure \ref{fig04} shows the normalized growth rate $\eta$($\equiv -\Im(\sigma)/\Re(\sigma)$ where $\Re(\sigma)$ and $\Im(\sigma)$ are the real and imaginary parts of the complex eigenfrequency)
for a template model (white square in Fig. \ref{evolution}) characterized by a stellar mass $M_{\star}= 0.3419 M_{\odot}$, $T_{\rm eff} = 31\, 089$K and a metallicity of $Z=0.05$. Positive (negative) values of $\eta$ implies instability (stability) for a particular mode. We consider radial modes ($\ell= 0$, blue dots) and also nonradial dipole ($\ell= 1$, black dots) and quadrupole 
($\ell= 2$, red dots) $g$ and $p$ modes. Note that, for our template model, we find unstable modes for the three harmonic degrees within the period range observed in the BLAPs. $g-$ modes with $\ell= 1, 2$ are unstable for radial order in the range $29 \leq k \leq 39$ and $54 \leq k \leq 67$, respectively. The radial fundamental mode ($r_0$) is also unstable, demonstrating that the modes observed in BLAPs with periods of $1200-1500$ s could be radial modes. We conclude that the observed periods in the BLAP stars can be explained by non--radial $g$ modes with high radial order or, alternatively, by low--order radial modes in the case of the shortest periods.

\begin{figure}
\includegraphics[width=\columnwidth]{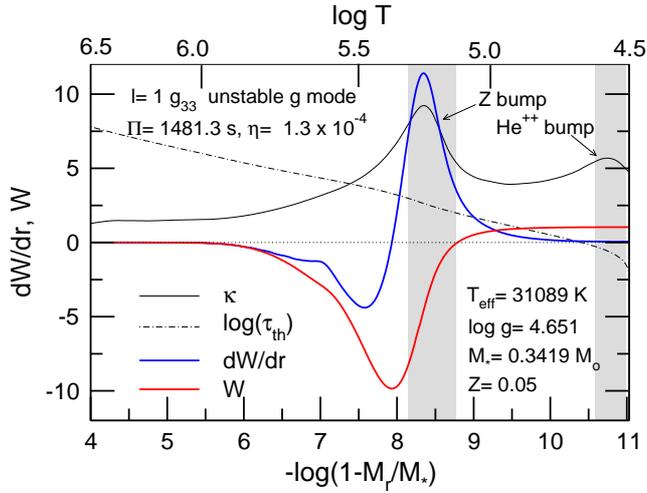}
\caption{Differential work ($dW/dr$) and the running work integral
($W$) in arbitrary units, in terms of the outer mass fraction
coordinate (lower scale) and the logarithm of the temperature (upper scale)
corresponding to the unstable $g$ mode with $\ell= 1$ and $k= 33$ (see
Fig. 4), along with the Rosseland opacity profile ($\kappa$) and the
logarithm of the thermal timescale ($\tau_{\rm th}$) of our $0.3419 M_{\sun}$ pre-WD template model ($T_{\rm eff}= 31100$ K).  The gray areas show the locations of two convection zones. Clearly visible is the $Z$ bump
in the opacity at $\log T \sim 5.25$, which is responsible for the
driving of the mode, while the He$^{++}$ bump ($\log T \sim 4.6$)
has no destabilizing effect.}
\label{fig05}
\end{figure}

To investigate the details of the driving/damping processes, we selected a representative unstable mode of our template model. Specifically, we chose the dipole $g-$mode with $k= 33$ (see Fig. \ref{fig04}). In Fig. \ref{fig05} we display the differential work ($dW/dr$) and the running work integral ($W$) in terms of the outer mass fraction coordinate (lower scale) and the logarithm of the local temperature (upper scale) for the selected unstable mode. Also, we depict the Rosseland opacity profile ($\kappa$) and the logarithm of the thermal timescale ($\tau_{\rm th}$). The region that destabilizes the mode (where $dW/dr >0$) is associated the the $Z-$bump in the opacity at $\log T \sim 5.25$, at the inner convection zone. Note that the bump in the opacity associated with the ionization of He (He$^{++}$ bump) has no destabilizing effects, as it does in the case of pre-ELMV stars \citep{2016A&A...588A..74C}. Indeed, the fact that the modes in BLAPs are destabilized at higher effective temperatures than in pre-ELMVs responds to the fact that the excitation in the former stars is due to the $Z$ bump, while in the pre-ELMVs the driving is due to the bump of HeII. 
Finally \citet{2018arXiv180204405P} reported a negative result in the search for BLAPs in the Magellanic System, implying that low metallicity stars are not pulsationally unstable. This is in line with a $Z$-bump excitation mechanism. 

\section{Conclusions}\label{conclusion}

We found that the position of the known BLAPs in the 
$\log T_{\rm eff}-\log g$ plane can be nicely explained by evolutionary sequences of pre-ELM WD stars with stellar masses between $0.27$ and $0.37 M_{\sun}$ and effective temperatures $T_{\rm eff} \sim 30\,000$ K. In addition, the stellar mass range is in perfect agreement with the values obtained from spectroscopy by \citet{2017NatAs...1E.166P}. Then, we conclude that the Blue Large-Amplitude Pulsator stars are in fact massive ($\sim 0.34-0.37 M_{\sun}$) pre-ELM WDs stars that are found in the hot stage induced by the H-shell burning.

From our analysis of the adiabatic pulsation properties we find that the observed periods and the rates of period change in the BLAP stars are consistent with high--order $g-$mode pulsations, with radial order in the range $ 25 \lesssim k  \lesssim 50$ for modes with $\ell= 1$ and $54 \lesssim k \lesssim 67$ for modes with $\ell= 2$. In particular, periods in the range $1200-1500$ s can also be due to low--order radial ($\ell= 0$) modes, including the fundamental ($k= 0$) radial mode. Finally, non-adiabatic computations proved that modes in this period range are unstable due to the $\kappa$ mechanism associated to a $Z-$bump in the opacity, as long as the metallicity $Z$ is substantially enhanced as compared with the solar metallicity. A high-metallicity evolutionary scenario will be explored in the future.  

\section*{Acknowledgements}

We thank the anonymous referee for her/his valuable comments and suggestions. Partial financial support from this research comes from CNPq and PRONEX-FAPERGS/CNPq (Brazil). Part of this work was supported by AGENCIA through the Programa de Modernizaci\'on Tecnol\'ogica BID 1728/OC-AR and the PIP 112-200801-00940 grant from CONICET. This research has made use of NASA Astrophysics Data System.

\label{lastpage}
\end{document}